\documentclass{article}

\usepackage{PRIMEarxiv}

\usepackage[utf8]{inputenc} 
\usepackage[T1]{fontenc}    
\usepackage{hyperref}       
\usepackage{url}            
\usepackage{booktabs}       
\usepackage{amsfonts}       
\usepackage{amsmath}
\usepackage{nicefrac}       
\usepackage{microtype}      
\usepackage{lipsum}
\usepackage{fancyhdr}  
\usepackage{caption}
\usepackage{subcaption}
\usepackage{graphicx}       
\graphicspath{{media/}}     
\usepackage{graphicx}
\usepackage[shortlabels]{enumitem}
\usepackage{enumerate}
\usepackage{booktabs,array}
\usepackage{multirow}
\usepackage{multicol}
\usepackage{makecell}

\usepackage{mathtools, cuted}

\usepackage{booktabs}

\usepackage{cite}

\usepackage{pgfplots}

\usepackage{subcaption}

\usepackage{balance}

\usepackage{tikz}
\usetikzlibrary{shapes.geometric, arrows, calc, fit,positioning}

\usepackage{caption}

\usepackage{enumitem}
 
\pgfplotsset{compat = newest}

\usepackage{comment}
\title{Delay analysis of the IEEE 802.11bd EDCA\\ with repetitions
\thanks{Aditya Sarkar is with the Department of Electrical and Computer Engineering, University of California San Diego, La Jolla, San Diego County, California, United States, 92037 (e-mail: adsarkar@ucsd.edu).\\
Sreelakshmi Manjunath is with the School of Computing and Electrical Engineering, Indian Institute of Technology Mandi, HP, India 175005 (e-mail: sreelakshmi@iitmandi.ac.in).\\
The work was done at Indian Institute of Technology Mandi.}}

\author{Aditya Sarkar\\
Department of Electrical and Computer Engineering\\
University of California San Diego\\
La Jolla, CA, United States\\
\And 
Sreelakshmi Manjunath\\
School of Computing and Electrical Engineering\\
Indian Institute of Technology Mandi\\
HP, India}

\begin{document}
\maketitle

\begin{abstract}
We analyse the performance of the IEEE 802.11bd MAC protocol, with Enhanced Distributed Channel Access (EDCA) and repeated transmissions, in terms of the MAC access delay of packets pertaining to safety-related events. We outline Markov chain models for the contention mechanism of priority-based access categories, and derive the associated steady-state probabilities. Using these probabilities, we characterise the delay experienced by the packet in the MAC layer. Further, we characterise the reliability of the protocol in terms of the likelihood that a packet is delivered within a critical time interval. Numerical computations are conducted to understand the impact of various system parameters on the MAC access delay. The analysis indicates that the MAC access delay depends on various system parameters, some of which are influenced by the traffic scenario and nature of safety-critical events. Motivated by this, we used our analysis to study the delay and reliability of the 802.11bd MAC protocol specific to the context of platooning of connected vehicles subject to interruptions by human-driven motorised two wheelers. We observe that while the delay performance of the protocol is as per the QoS requirements of the standard, the protocol may not be reliable for this specific application. Our study suggests that it is desirable to co-design vehicular communication protocols with prevalent safety-related traffic applications. 
\end{abstract}


\section{Introduction}
Vehicular communications is a promising technology for performing cooperative movement of connected cars, thereby improving the overall travelling experience. Specifically, road travels are much more safer, quality of ride is better and ride times are much shorter. Further, the cars can communicate with each other as it can send data and safety critical messages wirelessly. There is a multitude of literature \cite{sepulcre2021analytical, klapevz2021experimental, wang2016wideband, zheng2015performance, ren2022performance} that has analyzed the IEEE 802.11p EDCA performance of traffic scenario where all vehicles are assumed to move in a platoon and are connected by IEEE 802.11p. Further, with an aim to improve reliability, reduce delays and interference in vehicular communication, IEEE WiFi Task Group introduced a new protocol named IEEE 802.11bd.\\
\\
There are a number of improvements in the new protocol \cite{naik2019ieee, anwar2019physical, triwinarko2021phy, dingsurvey, souza2021effective}. Out of them, two crucial novelties that increases reliability are as follows. First, it is able to perform adaptive blind re-transmissions where it can send up to three repetitions of a packet with a time gap of SIFS (Short Interframe Space) between two. Second, the standard introduces channel bonding techniques in which the main channel (40MHz) is divided into two sub-channels of 20MHz. However, a station that uses 802.11bd can use either Enhanced Distributed Channel Access (EDCA) or a channel bonding. Recent research works have studied IEEE 802.11bd features. For instance, \cite{torgunakov2021study, torgunakov2022study, yacheur2021implementation} compared the performance of the channel bonding techniques with a single-channel access method of IEEE 802.11p, and demonstrates that the former is better than the latter in terms of quality of service. \cite{ma2021sinr} proposed an effective SINR-based model to conduct the QoS analysis for IEEE 802.11bd. Work done by \cite{zhuofei2023adaptive} introduced two new strategies for adaptive repetitions and showed its benefits over the current strategy. To the best of our knowledge, there is no work that analyzes the back-off procedure of adaptive repetitions technique for EDCA in terms of collision probability.\\
\\
This article presents a 2D Markov chain model for analyzing the EDCA mechanism of IEEE 802.11bd which is slightly different from IEEE 802.11p due to its ability to send up to 3 repetitions. In this study, we first analyze the EDCA mechanism with repetitions in terms of various system parameters. We also derive an expression for the reliability of the MAC protocol in terms of the probability that the delay is below a critical value. Through this we observe that the delay experienced by a data packet is dependent on parameters that are dependent on the traffic settings on the road. As QoS requirements of vehicular communication protocols is often specific to the application in question, we analyse the MAC access delay and the reliability of the MAC protocol for a specific case of platooning of connected vehicles in the presence of conventional human-driven vehicles.\\
\\ 
Overall, our contributions are two-fold and are summarized as follows:
\begin{itemize}
    \item First, we analyze the performance of backoff procedure of EDCA mechanism with repetitions using 2D Markov chain model for each access category. We further obtain the mean MAC access delay and standard deviation. We use numerical computations to understand the dependence of the delay characteristics on various network and protocol parameters. 
    \item We outline an exponential distribution-based model for  reliability of the MAC protocol. 
    \item Lastly, we illustrate our analysis for the specific application of platooning of connected vehicles subject to interruptions by human-driven motorised two wheelers. 
\end{itemize}

The rest of the paper is organized as follows. Section 2 presents the Markov chain analysis of IEEE 802.11bd EDCA with repetitions. In Section 3 we contrast our analysis with that of 802.11p. Section 4 presents an application of this analysis to platooning in a heterogeneous traffic setting. Lastly, Section 5 concludes our work.

\section{Mathematical Modeling and Analysis}
\subsection{Overview of IEEE 802.11bd MAC layer}
A station that uses IEEE 802.11bd standard in the MAC layer can utilize either channel bonding (CB) or an Enhanced Distributed Channel Access (EDCA) techniques with repetitions~\cite{zhuofei2023adaptive}. If a station uses channel bonding technique, then the channel is divided into two sub-channels - primary and secondary. If it is a channel bonding technique without fallback, then both the sub-channels are used for sending a packet. On the other hand, in case of channel bonding with fallback, the station will send the packet as soon as the primary sub-channel is idle. A station can also use EDCA with repetitions. Here, the EDCA mechanism at the MAC layer is the same as that of 802.11p~\cite{yao2019mac} except for the repetitions. Similar to 802.11p, it has four access categories AC0-AC3 where AC0 and AC1 are reserved for critical applications while the remaining two are for infotainment. The station accesses the channel through CSMA/CA and if it finds that it is idle for at least Arbitration Inter-Frame Space (AIFS) time, it sends not only the packet but also repetitions that follow the first packet with a time gap of Short Inter-Frame Space (SIFS). Further, SIFS is shorter than AIFS to ensure that the station does not lose its access to the channel.

The packet comprises of two parts - preamble and data. The data can be decoded only when the preamble is detected. If the receiver station fails to detect the preamble, then it is not aware of any arrival of data packet. Further, if it is able to decode the preamble, then it waits for the receiver station to decode the data. If it is able to decode the data, no further repetitions are sent. Otherwise, up to 3 more repetitions are sent. Overall, we have two sets of analyses. First, we will determine the stationary probabilities and using them, we calculate the mean and standard deviation for MAC delay.

\subsection{Markov Chain Model}
In this section, we outline a 2-D Markov chain (MC) model for the backoff behavior of the EDCA mechanism with repetitions. For the sake of this analysis,  we consider only two Access Categories (ACs), namely AC0 and AC1 in decreasing order of priority. The model is presented in Fig 1. The Markov chain model and the ensuing analysis is extendable to the two other low-priority categories AC2 and AC3, which are usually used for voice and video data. 

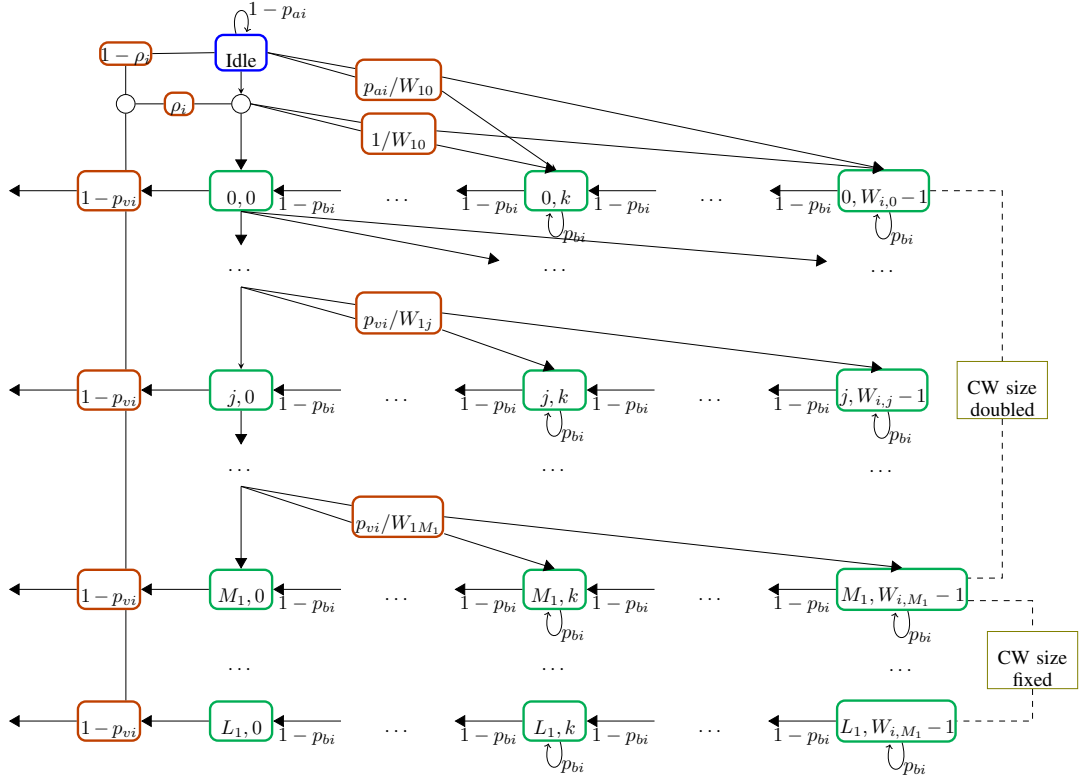
\begin{figure}[tbh]
\begin{center}
\scalebox{0.75}{
\begin{tikzpicture}
\definecolor{dkgreen}{rgb} {0.00,0.68,0.33}
\definecolor{dkred}{rgb}{0.75,0.25,0.0}
\definecolor{dkyellow}{rgb}{0.5,0.5,0.0}
\node[rectangle, rounded corners, draw=dkgreen, very thick, text width=1cm,minimum height=0.2cm,inner sep = 0.05cm] (00) at (0,0) {\begin{center}$0,0$\end{center}};
\node[rectangle, rounded corners, draw=dkred, very thick, text width=1cm, left=1.2cm of 00,minimum height=0.2cm,inner sep = 0.05cm] (Transmit1) {\begin{center}$1-p_{vi}$\end{center}};
\node[rectangle, rounded corners, draw=none, very thick, text width=2cm, left=1.2cm of Transmit1] (none0) {};
\node[rectangle, rounded corners, draw=none,very thick,  text width=2cm, right=1.2cm of 00, minimum height=0.005cm,inner sep = 0.0005cm] (cont01) {\begin{center}$\cdots$\end{center}};
\node[rectangle, rounded corners, draw=dkgreen,very thick,  text width=1cm, right=1.2cm of cont01,minimum height=0.2cm,inner sep = 0.05cm] (0k) {\begin{center}$0,k$\end{center}};
\node[rectangle, rounded corners, draw=none, text width=2cm, right=1.2cm of 0k,minimum height=0.005cm,inner sep = 0.0005cm] (cont02) {\begin{center}$\cdots$\end{center}};
\node[rectangle, rounded corners, draw=dkgreen,very thick,  text width=1.5cm,right=1.2cm of cont02,minimum height=0.2cm,inner sep = 0.05cm] (0W-1) {\begin{center} $0,W_{i,0}-1$\end{center}};
\node[rectangle, rounded corners, draw=none, text width=2cm, below=0.6cm of 00,minimum height=0.005cm,inner sep = 0.0005cm] (cont03) {\begin{center}$\cdots$\end{center}};
\node[rectangle, rounded corners, draw=none, text width=2cm, below=0.6cm of 0k,minimum height=0.005cm,inner sep = 0.0005cm] (cont04) {\begin{center}$\cdots$\end{center}};
\node[rectangle, rounded corners, draw=none, text width=2cm, below=0.6cm of 0W-1,minimum height=0.005cm,inner sep = 0.0005cm] (cont05) {\begin{center}$\cdots$\end{center}};
\draw[-triangle 60] (0W-1.west) -> (cont02.east)
	node[midway,below]{$1-p_{bi}$};
\draw[-triangle 60] (cont02.west) -> (0k.east)
	node[midway,below]{$1-p_{bi}$}; 
 \draw[-triangle 60] (0k.west) -> (cont01.east)
	node[midway,below]{$1-p_{bi}$};
\draw[-triangle 60] (cont01.west) -> (00.east)
	node[midway,below]{$1-p_{bi}$}; 
 \draw[-triangle 60] (00.west) -> node[below] {} (Transmit1.east); 
  \draw[-triangle 60] (Transmit1.west) -> node[below] {} (none0.east); 
  \draw[-triangle 60] (00.south) -> node[below] {} (cont03.north);
  \draw[-triangle 60] (00.south) -> node[below] {} (cont04.west);
  \draw[-triangle 60] (00.south) -> node[below] {} (cont05.west);
  \path (0k) edge [loop below] node [right] {$p_{bi}$} (0k);
  \path (0W-1) edge [loop below] node [right] {$p_{bi}$} (0W-1);
  \node[rectangle, rounded corners, draw=dkred,very thick,  text width=1.4cm,minimum height=0.2cm,inner sep = 0.05cm, below= 1.5cm of cont01] (Transition0j) {\begin{center}$p_{vi}/W_{1j}$\end{center}};
  \node[rectangle, rounded corners, draw=dkgreen, very thick, text width=1cm,minimum height=0.2cm,inner sep = 0.05cm, below= 2.8cm of 00] (j0) {\begin{center}$j,0$\end{center}};
\node[rectangle, rounded corners, draw=dkred,very thick, text width=1cm, left=1.2cm of j0,minimum height=0.2cm,inner sep = 0.05cm] (Transmitj) {\begin{center}$1-p_{vi}$\end{center}};
\node[rectangle, rounded corners, draw=none, text width=2cm, left=1.2cm of Transmitj] (nonej) {};
\node[rectangle, rounded corners, draw=none, text width=2cm, right=1.2cm of j0, minimum height=0.005cm,inner sep = 0.0005cm] (contj1) {\begin{center}$\cdots$\end{center}};
\node[rectangle, rounded corners, draw=dkgreen,very thick,  text width=1cm, right=1.2cm of contj1,minimum height=0.2cm,inner sep = 0.05cm] (jk) {\begin{center}$j,k$\end{center}};
\node[rectangle, rounded corners, draw=none, text width=2cm, right=1.2cm of jk,minimum height=0.005cm,inner sep = 0.0005cm] (contj2) {\begin{center}$\cdots$\end{center}};
\node[rectangle, rounded corners, draw=dkgreen,very thick,  text width=1.5cm,right=1.2cm of contj2,minimum height=0.2cm,inner sep = 0.05cm] (jW-1) {\begin{center} $j,W_{i,j}-1$\end{center}};
\node[rectangle, rounded corners, draw=none, text width=2cm, below=0.6cm of j0,minimum height=0.005cm,inner sep = 0.0005cm] (contj3) {\begin{center}$\cdots$\end{center}};
\node[rectangle, rounded corners, draw=none, text width=2cm, below=0.6cm of jk,minimum height=0.005cm,inner sep = 0.0005cm] (contj4) {\begin{center}$\cdots$\end{center}};
\node[rectangle, rounded corners, draw=none, text width=2cm, below=0.6cm of jW-1,minimum height=0.005cm,inner sep = 0.0005cm] (contj5) {\begin{center}$\cdots$\end{center}};
\draw[-triangle 60] (jW-1.west) -> (contj2.east)
	node[midway,below]{$1-p_{bi}$};
\draw[-triangle 60] (contj2.west) -> (jk.east)
	node[midway,below]{$1-p_{bi}$}; 
 \draw[-triangle 60] (jk.west) -> (contj1.east)
	node[midway,below]{$1-p_{bi}$};
\draw[-triangle 60] (contj1.west) -> (j0.east)
	node[midway,below]{$1-p_{bi}$}; 
 \draw[-triangle 60] (j0.west) -> node[below] {} (Transmitj.east); 
  \draw[-triangle 60] (Transmitj.west) -> node[below] {} (nonej.east); 
  \draw[-triangle 60] (j0.south) -> node[below] {} (contj3.north);
  \path (jk) edge [loop below] node [right] {$p_{bi}$} (jk);
  \path (jW-1) edge [loop below] node [right] {$p_{bi}$} (jW-1);
  \draw[-triangle 60] ([yshift=-0.3cm]Transition0j.east) -> node[below] {} (jk.north);
  \draw[-triangle 60] (Transition0j.east) -> node[below] {} (jW-1.north);
  \draw[-stealth] ([yshift=-0.2cm]cont03.south) -> node[below] {} (j0.north);
  \draw[-] ([yshift=-0.2cm]cont03.south) -- node[below] {} ([yshift=0.15cm]Transition0j.west);
  \draw[-] ([yshift=-0.2cm]cont03.south) -- node[below] {} ([yshift=-0.1cm]Transition0j.west);
    \node[rectangle, rounded corners, draw=dkred,very thick,  text width=1.5cm,minimum height=0.2cm,inner sep = 0.05cm, below= 1.6cm of contj1] (TransitionjM) {\begin{center}$p_{vi}/W_{1M_1}$\end{center}};
  \node[rectangle, rounded corners, draw=dkgreen,very thick,  text width=1cm,minimum height=0.2cm,inner sep = 0.05cm, below= 2.8cm of j0] (M0) {\begin{center}$M_1,0$\end{center}};
\node[rectangle, rounded corners, draw=dkred,very thick, text width=1cm, left=1.2cm of M0,minimum height=0.2cm,inner sep = 0.05cm] (TransmitM) {\begin{center}$1-p_{vi}$\end{center}};
\node[rectangle, rounded corners, draw=none, text width=2cm, left=1.2cm of TransmitM] (noneM) {};
\node[rectangle, rounded corners, draw=none, text width=2cm, right=1.2cm of M0, minimum height=0.005cm,inner sep = 0.0005cm] (contM1) {\begin{center}$\cdots$\end{center}};
\node[rectangle, rounded corners, draw=dkgreen, very thick, text width=1cm, right=1.2cm of contM1,minimum height=0.2cm,inner sep = 0.05cm] (Mk) {\begin{center}$M_1,k$\end{center}};
\node[rectangle, rounded corners, draw=none, text width=2cm, right=1.2cm of Mk,minimum height=0.005cm,inner sep = 0.0005cm] (contM2) {\begin{center}$\cdots$\end{center}};
\node[rectangle, rounded corners, draw=dkgreen, very thick, text width=2.2cm,right=1.2cm of contM2,minimum height=0.2cm,inner sep = 0.05cm] (MW-1) {\begin{center} $M_1,W_{i,M_1}-1$\end{center}};
\node[rectangle, rounded corners, draw=none, text width=2cm, below=0.6cm of M0,minimum height=0.005cm,inner sep = 0.0005cm] (contM3) {\begin{center}$\cdots$\end{center}};
\node[rectangle, rounded corners, draw=none, text width=2cm, below=0.6cm of Mk,minimum height=0.005cm,inner sep = 0.0005cm] (contM4) {\begin{center}$\cdots$\end{center}};
\node[rectangle, rounded corners, draw=none, text width=2cm, below=0.6cm of MW-1,minimum height=0.005cm,inner sep = 0.0005cm] (contM5) {\begin{center}$\cdots$\end{center}};
\draw[-triangle 60] (MW-1.west) -> (contM2.east)
	node[midway,below]{$1-p_{bi}$};
\draw[-triangle 60] (contM2.west) -> (Mk.east)
	node[midway,below]{$1-p_{bi}$}; 
 \draw[-triangle 60] (Mk.west) -> (contM1.east)
	node[midway,below]{$1-p_{bi}$};
\draw[-triangle 60] (contM1.west) ->  (M0.east)
	node[midway,below]{$1-p_{bi}$}; 
 \draw[-triangle 60] (M0.west) -> node[below] {} (TransmitM.east); 
  \draw[-triangle 60] (TransmitM.west) -> node[below] {} (noneM.east); 
  \path (Mk) edge [loop below] node [right] {$p_{bi}$} (Mk);
  \path (MW-1) edge [loop below] node [right] {$p_{bi}$} (MW-1);
  \draw[-triangle 60] ([yshift=-0.3cm]TransitionjM.east) -> node[below] {} (Mk.north);
  \draw[-triangle 60] (TransitionjM.east) -> node[below] {} (MW-1.north);
  \draw[-triangle 60] ([yshift=-0.2cm]contj3.south) -> node[below] {} (M0.north);
  \draw[-] ([yshift=-0.2cm]contj3.south) -- node[below] {} ([yshift=0.2cm]TransitionjM.west);
  \draw[-] ([yshift=-0.2cm]contj3.south) -- node[below] {} ([yshift=-0.1cm]TransitionjM.west);
  \node[rectangle, rounded corners, draw=dkgreen,very thick,  text width=1cm,minimum height=0.2cm,inner sep = 0.05cm, below= 1.6cm of M0] (L0) {\begin{center}$L_1,0$\end{center}};
\node[rectangle, rounded corners, draw=dkred,very thick, text width=1cm, left=1.2cm of L0,minimum height=0.2cm,inner sep = 0.05cm] (TransmitL) {\begin{center}$1-p_{vi}$\end{center}};
\node[rectangle, rounded corners, draw=none, text width=2cm, left=1.2cm of TransmitL] (noneL) {};
\node[rectangle, rounded corners, draw=none, text width=2cm, right=1.2cm of L0, minimum height=0.005cm,inner sep = 0.0005cm] (contL1) {\begin{center}$\cdots$\end{center}};
\node[rectangle, rounded corners, draw=dkgreen,very thick,  text width=1cm, right=1.2cm of contL1,minimum height=0.2cm,inner sep = 0.05cm] (Lk) {\begin{center}$L_1,k$\end{center}};
\node[rectangle, rounded corners, draw=none, text width=2cm, right=1.2cm of Lk,minimum height=0.005cm,inner sep = 0.0005cm] (contL2) {\begin{center}$\cdots$\end{center}};
\node[rectangle, rounded corners, draw=dkgreen,very thick,  text width=2cm,right=1.2cm of contL2,minimum height=0.2cm,inner sep = 0.05cm] (LW-1) {\begin{center} $L_1,W_{i,M_1}-1$\end{center}};
\draw[-triangle 60] (LW-1.west) ->  (contL2.east)
	node[midway,below]{$1-p_{bi}$};
\draw[-triangle 60] (contL2.west) ->  (Lk.east)
	node[midway,below]{$1-p_{bi}$}; 
 \draw[-triangle 60] (Lk.west) ->  (contL1.east)
	node[midway,below]{$1-p_{bi}$};
\draw[-triangle 60] (contL1.west) ->  (L0.east)
	node[midway,below]{$1-p_{bi}$}; 
 \path (Lk) edge [loop below] node [right] {$p_{bi}$} (Lk);
  \path (LW-1) edge [loop below] node [right] {$p_{bi}$} (LW-1);
  \draw[-triangle 60] (L0.west) -> node[below] {}(TransmitL.east); 
  \draw[-triangle 60] (TransmitL.west) -> node[below] {}(noneL.east); 
 \draw[-] ([xshift=0.3cm]Transmit1.south) -- node[below] {} ([xshift=0.3cm]Transmitj.north);
 \draw[-] ([xshift=0.3cm]Transmitj.south) -- node[below] {} ([xshift=0.3cm]TransmitM.north);
 \draw[-] ([xshift=0.3cm]TransmitM.south) -- node[below] {} ([xshift=0.3cm]TransmitL.north);
 \node[circle, draw, above=1cm of 00] (sum2) {};
  \node[circle, draw, left=1.7cm of sum2] (sum1) {};
  \draw[-] ([xshift=0.3cm]Transmit1.north) -- node[below] {} (sum1.south);
  \node[rectangle, rounded corners, draw=dkred, very thick, text width=0.5cm, right=0.5cm of sum1, minimum height=0.2cm,inner sep = 0.001cm] (saturated) {\begin{center}\vspace{-2mm}$\rho_i$\end{center}};
    \draw[-] (sum1.east) -- node[below] {} (saturated.west);
    \draw[-]  (saturated.east) -- node[below] {} (sum2.west) ;
   \draw[-triangle 60] (sum2.south) -> node[below] {} (00.north);  
\node[rectangle, rounded corners, draw=dkred,very thick,  text width=1.2cm,minimum height=0.2cm,inner sep = 0.05cm, above= 0.35cm of cont01] (Transition0) {\begin{center}$1/W_{10}$\end{center}};
  \draw[-] (sum2.east) -- node[below] {} ([yshift=0.2cm]Transition0.west);
  \draw[-] (sum2.east) -- node[below] {} ([yshift=0.05cm]Transition0.west);
  \draw[-triangle 60] ([yshift=-0.2cm]Transition0.east) -> node[below] {} (0k.north); 
  \draw[-triangle 60] ([yshift=0.05cm]Transition0.east) -> node[below] {} (0W-1.north);
  \node[rectangle, rounded corners, draw=dkred,very thick,  text width=0.9cm, above=0.5cm of sum1, minimum height=0.2cm,inner sep = 0.001cm] (notsaturated) {\begin{center}\vspace{-2mm}$1-\rho_i$\end{center}};
  \node[rectangle, rounded corners, draw=blue, very thick, text width=0.8cm,above=0.4cm of sum2,minimum height=0.2cm,inner sep = 0.05cm] (idle) {\begin{center}Idle\end{center}};
 \draw[-] (sum1.north) -- node[below] {} (notsaturated.south); 
 \draw[-] (idle.west) -- node[below] {} (notsaturated.east);
 \draw[-stealth] (idle.south) -> node[below] {} (sum2.north); 
 \path (idle) edge [loop above] node [right] {$1-p_{ai}$} (idle);
 \node[rectangle, rounded corners, draw=dkred,very thick,  text width=1.4cm,minimum height=0.2cm,inner sep = 0.05cm, above= 0.2cm of Transition0] (Transition0newentry) {\begin{center}$p_{ai}/W_{10}$\end{center}};
   \draw[-] (idle.east) -- node[below] {} ([yshift=0.2cm]Transition0newentry.west);
  \draw[-] (idle.east) -- node[below] {} ([yshift=0.05cm]Transition0newentry.west);
  \draw[-stealth] ([yshift=-0.2cm]Transition0newentry.east) -> node[below] {} (0k.north); 
  \draw[-triangle 60] ([yshift=0.05cm]Transition0newentry.east) -> node[below] {} (0W-1.north);
  \node[rectangle, draw=dkyellow, text width=1.5cm,minimum height=0.2cm,inner sep = 0.05cm, right= 0.5cm of jW-1] (Doubled) {\begin{center}CW size\\ doubled\end{center}};
  \node[rectangle, draw=dkyellow, text width=1.5cm,minimum height=0.2cm,inner sep = 0.05cm, right= 0.5cm of contM5] (Fixed) {\begin{center}CW size\\ fixed\end{center}};
  \draw[dashed] (0W-1.east) -| node[below] {} (Doubled.north);
  \draw[dashed] (Doubled.south) |- node[below] {}([yshift=0.2cm]MW-1.east);
    \draw[dashed] ([yshift=-0.2cm]MW-1.east) -| node[below] {} (Fixed.north);
    \draw[dashed] (Fixed.south) |- node[below] {} (LW-1.east); 
\end{tikzpicture}
}\end{center}
\caption{Markov chain for ACi.}
\label{overview}
\end{figure}

Similar to previous studies on analyzing IEEE 802.11p EDCA mechanism, we assume that the channel is ideal and the traffic is saturated~\cite{Gayathree802.11p}.

The backoff stage and counter are the two states considered for the 2-D Markov chain. At a given time instant $t$, they are represented by the $\{s(t), b(t)\}$. The contention window (CW) doubles at each backoff stage, and the maximum number of times (M) it can do so is given by
\begin{equation}
    M = \text{log}_{2} \frac{CW_{max} + 1}{CW_{min} + 1},\label{eq:CW_doubling_limit}
\end{equation}
where $CW_{min}$ and $CW_{max}$ denotes the minimum and maximum contention window sizes, respectively. Further, the CW for any $j^{th}$ backoff stage for i$^{th}$ AC is given by
\begin{equation}
    W_{i,j} = \begin{cases} 
      2^j W_{i,0} & j\leq M \\
      2^M W_{i,0} & M < j\leq L  
   \end{cases}\label{eq:CW_limit}
\end{equation}
where $W_{i,j}$ is the limit on $j^{th}$ backoff stage and $i^{th}$ AC. Here, $L$ represents the retry limit.

For AC i where i $\in$ \{0,1\}, we write the following state transitions for backoff procedure:
\begin{itemize}
    \item $\{j,k+1\} \rightarrow \{j,k\}$ - The backoff counter is decremented when the channel is idle, this occurs with probability
    \begin{equation}
        P(i,k|i,k+1) = 1 - p_{bi}. \label{eq:BC_reduceby1}
    \end{equation}
    \item $\{j,k\} \rightarrow\{j,k\} $ - The backoff counter is not decremented when the channel is busy, the probability of which is
    \begin{equation}
        P(j,k|j,k) = p_{bi}.\label{eq:BC_same}
    \end{equation}
    \item $\{j-1,0\} \rightarrow\{j,k\}$    - This transition occurs when the backoff stage is incremented by one, which implies that there has been an internal collision or that the preamble has not been detected at the receiver. The associated probability is given by
    \begin{equation}
        P(j,k|j-1,0) = (1-\rho_i) \left(\frac{p_{vi}}{W_{i,j}} + \frac{1-p_{vi}}{W_{i,j}} p_{o} \right),\label{eq:BC_increment}
    \end{equation} 
    where $p_{vi}$ is the probability of internal collision and $p_{o}$ is the probability of external collision given by
    \begin{equation}
        p_{o} = \sum_{i=1}^4 p(Z=i)p_{oi},\label{eq:external_collision}
    \end{equation}
     where $p_{oi}$ is the probability that there is an external collision when $i$ packets are sent and is defined as 
     \begin{align*}
         p_{o1} &= p_{ex}, \notag \\  
         p_{o2} &= 2p_{ex} - p^2_{ex}, \notag \\  
         p_{o3} &= 3p_{ex} - 3p^2_{ex} + p^3_{ex}, \notag \\  
         p_{o4} &= 4p_{ex} - 6p^2_{ex} + 4p^{3}_{ex} - p^4_{ex},  
     \end{align*}
    where $p_{ex} = 1 - (1-(\beta_0 + \beta_1))^{Ncs-1}$, $\beta_i$ is external transmission probability of AC i and $N_{cs}$ is the number of stations. Further $Z$ is a random variable that denotes the number of packets sent and is given as
    \begin{align}
        p(Z=1) =&\, p_d p_s \notag \\
        p(Z=2) =&\, p_d(1-p_s)p_sp_d + (1-p_d)p_sp_d \notag \\
        p(Z=3) =&\, p_d(1-ps)p_d(1-p_s)p_sp_d + p_d(1-p_s)(1-p_d)p_dp_s + (1-p_d)p_d(1-p_s)p_sp_d \notag\\
        &+ (1-p_d)(1-p_d)p_dp_s \notag \\
        p(Z=4) =&\, 1 - \sum_{i=1}^3 p(Z=i).\label{eq:prob_Z}
    \end{align}
Note that $\sum_{i=1}^4 p(Z = i) = 1$. Here $p_d$ is the probability that preamble is detected and $p_s$ is the probability that data is decoded successfully.
    \item $\{j,0\} \rightarrow \{0,k\}$ - This happens when the data is detected in either in the first transmission or during the repetitions, the probability of which is
    \begin{equation}
        P(0,k|j,0) = \rho_i \left( \frac{1-p_{vi}}{W_{i,j}} (1-p_{o}) \right).\label{eq:prob_Z0}
    \end{equation}
\end{itemize}
Equating the sum of steady-state probabilities for all states in Markov chain to 1, we obtain the internal transmission probability for AC0 and AC1 i.e. $\omega_{0}$ and $\omega_{1}$ as 
\begin{align}
    \omega_0 =&\, \left[\frac{(W_{0,0} + 1)}{2(1-p_{b0})} + \frac{1-\rho_0}{p_{a0}}\right]^{-1},\label{eq:omega0}\\
    \omega_1 =&\, \frac{1-p_{c1}^{L+1}}{1-p_{c1}}\bigg [ \frac{1-p_{c1}^{L+1}}{1-p_{c1}} + \frac{W_{1,0} - 1}{2(1-p_{b1})} + \bigg(\frac{1}{2(1-p_{b1})} \frac{W_{1,0} 2p_{c1} (1-(2p_{c1})^M)}{1-2p_{c1}} - \frac{p_{c1} (1-p_{c1}^M)}{1-p_{c1}} \notag\\
    &+ \frac{(2^M W_{1,0} - 1)(1-p_{c1}^{L-M})p_{c1}^{M+1}}{1-p_{c1}}\bigg) + \frac{1-\rho_1}{p_{a1}}\bigg ]^{-1},\label{eq:omega1}
\end{align}
where $p_{c1}$ is the probability that an AC1 packet undergoes a collision, and is given by 
\begin{equation}
    p_{c1} = p_{v1} + (1-p_{v1})p_{o}.\label{eq:AC1_collision_probability}
\end{equation}
The probability of internal collision p$_{vi}$ for each AC is given by 
\begin{align}
    p_{v0} &= 0, \notag \\
    p_{v1} &= \omega_0 \label{eq:internal_collision_probabilities}
\end{align}
Furthermore, the external transmission probabilities are then obtained as
\begin{align}
    \beta_0 &= \omega_0, \notag \\
    \beta_1 &= \omega_1(1-\omega_0).\label{eq:external_transmission_probabilities}
\end{align}
Having derived all the probabilities, we have the complete Markov chain model for the two ACs under study. We now proceed to use this model to characterise the time delay experienced by a data packet in the MAC layer. 
\subsection{Delay Analysis}
The average transmission time can be calculated as follows
\begin{equation}
    T_{tr} = \frac{PHY_H}{R_b} + \frac{MAC_H + \mathbb{E}[P]}{R_d} + \delta, \label{eq:transmission_time}
\end{equation}
where PHY$_H$ and MAC$_H$ are header lengths, $\delta$ is the propagation delay, R$_b$ is the basic rate, R$_d$ is the data rate and $\mathbb{E}[P]$. Given that a packet could be transmitted upto 4 times, the mean packet size can be derived as
\begin{align}
    \mathbb{E}[P] = p(Z=1) P + p(Z=2) 2P + p(Z=3) 3P + p(Z=4) 4P \label{eq:mean_packet_size}
\end{align}  
where $p(Z = i)$ is defined as in equation~\eqref{eq:prob_Z}.

Let $q_{i,k}$ denote the steady-state probability that the service time of packet, \emph{i.e.} the MAC access delay experienced by the packet, is equal to t$_{s_i,k}$. The corresponding Probability Generating Function (PGF) can be written as
\begin{equation}
    P_{T_{si}}(z) = \sum_{k=0}^{\infty} q_{i,k} z^{t_{s_{i,k}}}. \label{eq:PGF}
\end{equation}
The service time comprises of backoff and transmission time, and both of these are random variables. Since the transmission time depends on the number of repetition required for successful detection at the receiver, the PGF for transmission time (T$_{tr}$) is given by
\begin{align}
    TR_i(z) =&\, p(Z=1) z^{T_{tr}} + p(Z=2) z^{2 T_{tr} + SIFS[i]} + p(Z=3) z^{3 T_{tr} + 2 SIFS[i]} + p(Z=4) z^{4 T_{tr} + 3 SIFS[i]}, \label{eq:PGF_TR}
\end{align}
where SIFS[i] is the Short Inter Frame Space for ACi. Similarly, the PGF of the time 
that the backoff counter of ACi decrements by one is given as 
\begin{align}
H_{i}(z) =&\, (1-p_{bi})z^{\sigma} + p_{bi}\bigg(p(Z=1) z^{T_{tr} + AIFS[i]} 
+ p(Z=2) z^{2 T_{tr} + SIFS[i] + AIFS[i]}\notag  \\ 
&+ p(Z=3)z^{3 T_{tr} + 2 SIFS[i] +  AIFS[i]}
+ p(Z=4) z^{4 T_{tr} + 3 SIFS[i] + AIFS[i]}\bigg). \label{eq:PGF_backoff}
\end{align}
For AC0 and AC1, we can now write the functions P$_{T_{s0}}$(z) and P$_{T_{s1}}$(z) as
\begin{align}
    P_{T_{s0}}(z) &=\, \frac{TR(z)}{W_{0,0}} \displaystyle{\sum_{k=0}^{W_{0,0}-1}} [H_{0}(z)]^k, \label{eq:PGF_AC0} \\ 
    P_{T_{s1}}(z) &=\, (1-p_{c1}) TR_1(z) \displaystyle{\sum_{n=0}^{L}} p^n_{v1} \pi_{j=0}^{n} B_{1,j}(z) + p_{c1}^{L+1} \pi_{j=0}^{L} B_{1,j}(z)\,\,\,
    \text{ where }\label{eq:PGF_AC1} \\
    B_{1,j}(z) &=\, \begin{cases}\frac{1}{W_{1j}} \displaystyle{\sum_{k=0}^{W_{1j} - 1}} [H_1(z)]^k, \text{  } j \in (1,M) \notag \vspace{2mm}\\
     \frac{1}{W_{1M}} \displaystyle{\sum_{k=0}^{W_{1M} - 1} }[H_1(z)]^k, \text{  } j \in (M,L) \end{cases} \notag
\end{align}
Using the PGFs derived above, we compute the mean and standard deviation of the MAC access delay as follows
\begin{align}
    T_{Si} &=\, P_{T_{si}}^{'}(1) \label{eq:mean_delay} \\
    D_{T_{Si}} &= P_{T_{si}}^{''}(1) + P_{T_{si}}^{'}(1) - (P_{T_{Si}}(1))^2, \,\,\,i={0,1}  \label{eq:stdev_delay}
\end{align}
where $P_{T_{si}}^{'}(1)$ denotes the first derivative of the PGF evaluated at $1$. 

\subsection{Reliability analysis}
To compute the reliability of the protocol, we obtained the PMF from PGF as follows -  
\begin{equation}
    P(t_{si} = k) = \frac{1}{k!} \frac{\partial^k P_{Tsi}(z)}{\partial z^k}
\end{equation}
Similar to work done in \cite{yao2019mac}, we assume that the PMF follows a shifted exponential distribution as follows - 
\begin{equation}
    P(t_{si} = k) = \begin{cases} 
      \theta_i e^{-\theta_i(x - T_{tr})} & x\geq T_{tr} \\
      0 & x < T_{tr} 
   \end{cases}
\end{equation}
where $\theta_i^{-1} = \sqrt{\mathbb{E}(t^2_{si}) - \mathbb{E}(t_{si})^2}$. Using this PMF, we can compute the reliability $\mathcal{R}$ as follows - 
\begin{align}
    \mathcal{R} &= P(t_{si} \leq \tau)\notag \\
                &= \begin{cases} 
      1 - e^{-\theta_i(\tau - T_{tr})}  & \tau \geq T_{tr} \\
      0 & \tau < T_{tr} \label{eq:reliability}
   \end{cases}               
\end{align}
for $\tau \geq 0$.\\

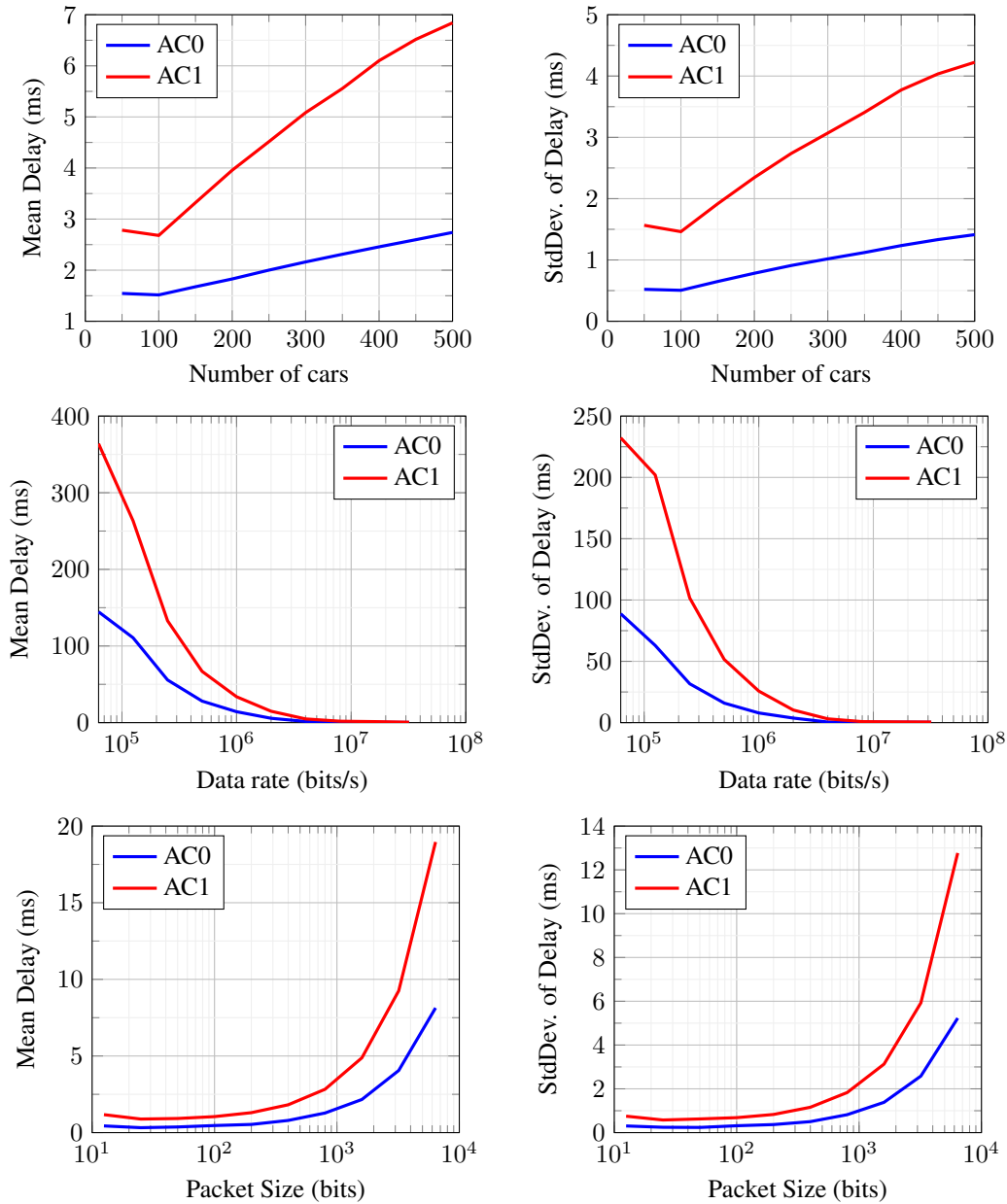
\begin{figure}[tbh]
\centering
\begin{tabular}[c]{c c}
\begin{tikzpicture}
\begin{axis}[
yticklabel style={
/pgf/number format/fixed,
/pgf/number format/precision=2,
/pgf/number format/fixed zerofill,
/pgf/number format/fixed relative,
},
    every axis plot/.append style={thick},
    xlabel = {Number of cars},
    ylabel = {Mean Delay (ms)},
    xmin = 0, xmax = 500,
    ymin = 1, ymax = 7,
    xtick distance = 100,
    ytick distance = 1,
    grid = both,
    minor tick num = 1,
    major grid style = {lightgray},
    minor grid style = {lightgray!25},
    width = 0.4\textwidth,
    height = 0.35\textwidth,
    legend cell align = {right},
    legend pos = north west
]
 
\addplot[blue,style=very thick]  table [x = {x}, y = {y}] {delay_cars.prn};
\addplot[red,style=very thick]  table [x = {x}, y = {a}] {delay_cars.prn};
\legend{
AC0 , 
AC1 }
\end{axis}
\end{tikzpicture}
&
\begin{tikzpicture}
\begin{axis}[
yticklabel style={
/pgf/number format/fixed,
/pgf/number format/precision=2,
/pgf/number format/fixed zerofill,
/pgf/number format/fixed relative,
},
    every axis plot/.append style={thick},
    xlabel = {Number of cars},
    ylabel = {StdDev. of Delay (ms)},
    xmin = 0, xmax = 500,
    ymin = 0, ymax = 5,
    xtick distance = 100,
    ytick distance =1,
    grid = both,
    minor tick num = 1,
    major grid style = {lightgray},
    minor grid style = {lightgray!25},
    width = 0.4\textwidth,
    height = 0.35\textwidth,
    legend cell align = {right},
    legend pos = north west
]
 
\addplot[blue,style=very thick]  table [x = {x}, y = {b}] {delay_cars.prn};
\addplot[red,style=very thick]  table [x = {x}, y = {z}] {delay_cars.prn};
\legend{
AC0 , 
AC1 }
\end{axis}
\end{tikzpicture}
\\
\begin{tikzpicture}
\begin{semilogxaxis}[
yticklabel style={
/pgf/number format/fixed,
/pgf/number format/precision=2,
/pgf/number format/fixed zerofill,
/pgf/number format/fixed relative,
},
    every axis plot/.append style={thick},
    xlabel = {Data rate (bits/s)},
    ylabel = {Mean Delay (ms)},
    xmin = 62500, xmax = 100000000,
    ymin = 0, ymax = 400,
    ytick distance = 100,
    grid = both,
    minor tick num = 1,
    major grid style = {lightgray},
    minor grid style = {lightgray!25},
    width = 0.4\textwidth,
    height = 0.35\textwidth,
    legend cell align = {right},
    legend pos = north east
]
 
\addplot[blue,style=very thick]  table [x = {x}, y = {y}] {delay_data.prn};
\addplot[red,style=very thick]  table [x = {x}, y = {a}] {delay_data.prn};
\legend{
AC0 , 
AC1 }
\end{semilogxaxis}
\end{tikzpicture}
& 
\begin{tikzpicture}
\begin{semilogxaxis}[
yticklabel style={
/pgf/number format/fixed,
/pgf/number format/precision=2,
/pgf/number format/fixed zerofill,
/pgf/number format/fixed relative,
},
    every axis plot/.append style={thick},
    xlabel = {Data rate (bits/s)},
    ylabel = {StdDev. of Delay (ms)},
    xmin = 62500, xmax = 100000000,
    ymin = 0, ymax = 250,
    ytick distance = 50,
    grid = both,
    minor tick num = 1,
    major grid style = {lightgray},
    minor grid style = {lightgray!25},
    width = 0.4\textwidth,
    height = 0.35\textwidth,
    legend cell align = {right},
    legend pos = north east
]
 
\addplot[blue,style=very thick]  table [x = {x}, y = {z}] {delay_data.prn};
\addplot[red,style=very thick]  table [x = {x}, y = {b}] {delay_data.prn};
\legend{
AC0 , 
AC1 }
\end{semilogxaxis}
\end{tikzpicture}
\\
\begin{tikzpicture}
\begin{semilogxaxis}[
yticklabel style={
/pgf/number format/fixed,
/pgf/number format/precision=2,
/pgf/number format/fixed zerofill,
/pgf/number format/fixed relative,
},
    every axis plot/.append style={thick},
    xlabel = {Packet Size (bits)},
    ylabel = {Mean Delay (ms)},
    xmin = 10, xmax = 10000,
    ymin = 0, ymax = 20,
    ytick distance = 5,
    grid = both,
    minor tick num = 1,
    major grid style = {lightgray},
    minor grid style = {lightgray!25},
    width = 0.4\textwidth,
    height = 0.35\textwidth,
    legend cell align = {right},
    legend pos = north west
]
 
\addplot[blue,style=very thick]  table [x = {x}, y = {y}] {delay_pktsize.prn};
\addplot[red,style=very thick]  table [x = {x}, y = {a}] {delay_pktsize.prn};
\legend{
AC0 , 
AC1 }
\end{semilogxaxis}
\end{tikzpicture}
&
\begin{tikzpicture}
\begin{semilogxaxis}[
yticklabel style={
/pgf/number format/fixed,
/pgf/number format/precision=2,
/pgf/number format/fixed zerofill,
/pgf/number format/fixed relative,
},
    every axis plot/.append style={thick},
    xlabel = {Packet Size (bits)},
    ylabel = {StdDev. of Delay (ms)},
    xmin = 10, xmax = 10000,
    ymin = 0, ymax = 14,
    ytick distance = 2,
    grid = both,
    minor tick num = 1,
    major grid style = {lightgray},
    minor grid style = {lightgray!25},
    width = 0.4\textwidth,
    height = 0.35\textwidth,
    legend cell align = {right},
    legend pos = north west
]
 
\addplot[blue,style=very thick]  table [x = {x}, y = {z}] {delay_pktsize.prn};
\addplot[red,style=very thick]  table [x = {x}, y = {b}] {delay_pktsize.prn};
\legend{
AC0 , 
AC1}
\end{semilogxaxis}
\end{tikzpicture}
\end{tabular}
\caption{Delay characteristics as a function of network parameters (i) number of transmitting vehicles, (ii) data rate, (iii) packet size. The mean delay is seen to increase with an increase in the number of transmitting vehicle, and as the packet size increases.}
\label{fig:Mean_StDev_General1}
\end{figure}

\subsection{Computations}
We have now derived all the quantities required to characterise the MAC access delay. While deriving closed-form expressions for the mean and standard deviation of the MAC access delay is rather arduous, one may use the Markov chain model and PGFs to compute these values for a given set of parameter values. We now illustrate the same.

For the sake of this illustration, we consider two types of safety-related data packets: (i) event-driven safety-critical messages, and (ii) routine safety messages. The packets pertaining to the safety-critical messages are assigned the high priority AC0 category, and those corresponding to routine safety messages use AC1. Suppose packets arrive at AC0 MAC queue as per a Poisson process with mean rates $\lambda_0$ pkts/sec, and at the AC1 MAC queue periodically at rate $\lambda_1$ pkts/sec. The arrival probabilities for a time interval of $\epsilon$ are then given by 
\begin{align*}
    p_{a0} =\, \sum_{k=1}^{\infty} \frac{(\lambda_0 \epsilon)^k}{k!} = 1 - e^{\lambda_0 \epsilon},  &&
     p_{a1} =\, \lambda_1 \epsilon.
\end{align*}
The rates $\lambda_0$ and $\lambda_1$ may be fixed as per the traffic scenario being evaluated. For a given set of parameter values, we first consider an initial guess for $\rho_0, \rho_1$ and then compute $\omega_0$ and $\omega_1$ by numerically solving non-linear equations ~\eqref{eq:omega0} and ~\eqref{eq:omega1}. Next, we compute the mean delays T$_{Si}$ using ~\eqref{eq:mean_delay} and standard deviations of delay using ~\eqref{eq:stdev_delay}. Using the mean, we calculate the utilization factor as $\rho_i$ = $\lambda_i T_{Si}$. This is done multiple times until $\rho_i$ is close to our initial guess within a threshold of 10$^{-5}$. Probabilities for preamble detection and data decoding are taken as $0.9$ and $0.8$ respectively. We have assumed a fixed value of these probabilities as these dependent on various PHY attributes, which is beyond the scope of this work. Note that for $p_d = p_s = 1, p(Z=1) = 1$ and $p(Z=i) = 0$ for $i\in \{2,3,4\}$, which is same case as in 802.11p, the predecessor of 802.11bd. 
\begin{table}[hbt!]
\centering
\begin{tabular}{|l|l|l|l|}
\hline
\textbf{Parameter}           & \textbf{Value} & \textbf{Parameter} & \textbf{Value} \\ \hline
Basic rate                   & 1 Mbps         & Data rate          & 27 Mbps        \\ \hline
PHY header                   & 48 bits        & Retry limit        & 2              \\ \hline
Slot time                    & 13 $\mu$s      & SIFS               & 32 $\mu$s      \\ \hline
Packet size (P)              & 4000 bits      & AIFS 0             & 2              \\ \hline
CW$_{1,min}$, CW$_{1,max}$ & 15, 31         & AIFS 1             & 3              \\ \hline
\end{tabular}
\caption{Network and protocol parameter values}
\label{Tab:Tcr1}
\end{table}

\begin{figure}[tbh]
\centering
\begin{tabular}[c]{c c}
\begin{tikzpicture}
\begin{axis}[
yticklabel style={
/pgf/number format/fixed,
/pgf/number format/precision=2,
/pgf/number format/fixed zerofill,
/pgf/number format/fixed relative,
},
    every axis plot/.append style={thick},
    xlabel = {AC0 pkt generation rate, $\lambda_0$ (pkts/sec)},
    ylabel = {Mean Delay (ms)},
    xmin = 10, xmax = 55,
    ymin = 0, ymax = 3,
    xtick distance = 10,
    ytick distance = 0.5,
    grid = both,
    minor tick num = 1,
    major grid style = {lightgray},
    minor grid style = {lightgray!25},
    width = 0.4\textwidth,
    height = 0.35\textwidth,
    legend cell align = {right},
    legend pos = north east
]
 
\addplot[blue,style=very thick]  table [x = {x}, y = {y}] {delay_lambda0.prn};
\addplot[red,style=very thick]  table [x = {x}, y = {a}] {delay_lambda0.prn};
\legend{
AC0 , 
AC1
}
\end{axis}
\end{tikzpicture}
&
\begin{tikzpicture}
\begin{axis}[
yticklabel style={
/pgf/number format/fixed,
/pgf/number format/precision=2,
/pgf/number format/fixed zerofill,
/pgf/number format/fixed relative,
},
    every axis plot/.append style={thick},
    xlabel = {AC0 pkt generation rate, $\lambda_0$ (pkts/sec)},
    ylabel = {StDev. of Delay (ms)},
    xmin = 10, xmax = 55,
    ymin = 0, ymax = 2,
    xtick distance = 10,
    ytick distance = 0.5,
    grid = both,
    minor tick num = 1,
    major grid style = {lightgray},
    minor grid style = {lightgray!25},
    width = 0.4\textwidth,
    height = 0.35\textwidth,
    legend cell align = {right},
    legend pos = north east
]
 
\addplot[blue,style=very thick]  table [x = {x}, y = {z}] {delay_lambda0.prn};
\addplot[red,style=very thick]  table [x = {x}, y = {b}] {delay_lambda0.prn};
\legend{
AC0 , 
AC1
}
\end{axis}
\end{tikzpicture}\\
\begin{tikzpicture}
\begin{axis}[
yticklabel style={
/pgf/number format/fixed,
/pgf/number format/precision=2,
/pgf/number format/fixed zerofill,
/pgf/number format/fixed relative,
},
    every axis plot/.append style={thick},
    xlabel = {AC1 pkt generation rate, $\lambda_1$ (pkts/sec)},
    ylabel = {Mean Delay (ms)},
    xmin = 10, xmax = 55,
    ymin = 0, ymax = 3,
    xtick distance = 10,
    ytick distance = 0.5,
    grid = both,
    minor tick num = 1,
    major grid style = {lightgray},
    minor grid style = {lightgray!25},
    width = 0.4\textwidth,
    height = 0.35\textwidth,
    legend cell align = {right},
    legend pos = north east
]
 
\addplot[blue,style=very thick]  table [x = {x}, y = {y}] {delay_lambda1.prn};
\addplot[red,style=very thick]  table [x = {x}, y = {a}] {delay_lambda1.prn};
\legend{
AC0 , 
AC1 ,}
\end{axis}
\end{tikzpicture}
&
\begin{tikzpicture}
\begin{axis}[
yticklabel style={
/pgf/number format/fixed,
/pgf/number format/precision=2,
/pgf/number format/fixed zerofill,
/pgf/number format/fixed relative,
},
    every axis plot/.append style={thick},
    xlabel = {AC1 pkt generation rate, $\lambda_1$ (pkts/sec)},
    ylabel = {StdDev. of Delay (ms)},
    xmin = 10, xmax = 55,
    ymin = 0, ymax = 2,
    xtick distance = 10,
    ytick distance = 0.5,
    grid = both,
    minor tick num = 1,
    major grid style = {lightgray},
    minor grid style = {lightgray!25},
    width = 0.4\textwidth,
    height = 0.35\textwidth,
    legend cell align = {right},
    legend pos = north east
]
 
\addplot[blue,style=very thick]  table [x = {x}, y = {z}] {delay_lambda1.prn};
\addplot[red,style=very thick]  table [x = {x}, y = {b}] {delay_lambda1.prn};
\legend{
AC0, 
AC1}
\end{axis}
\end{tikzpicture}
\end{tabular}
\caption{Characteristics of the MAC access delay as a function of the mean rate of packet generation. The delay is seen to increase as the mean rate increases.}
\label{fig:Mean_StDev_General2}
\end{figure}
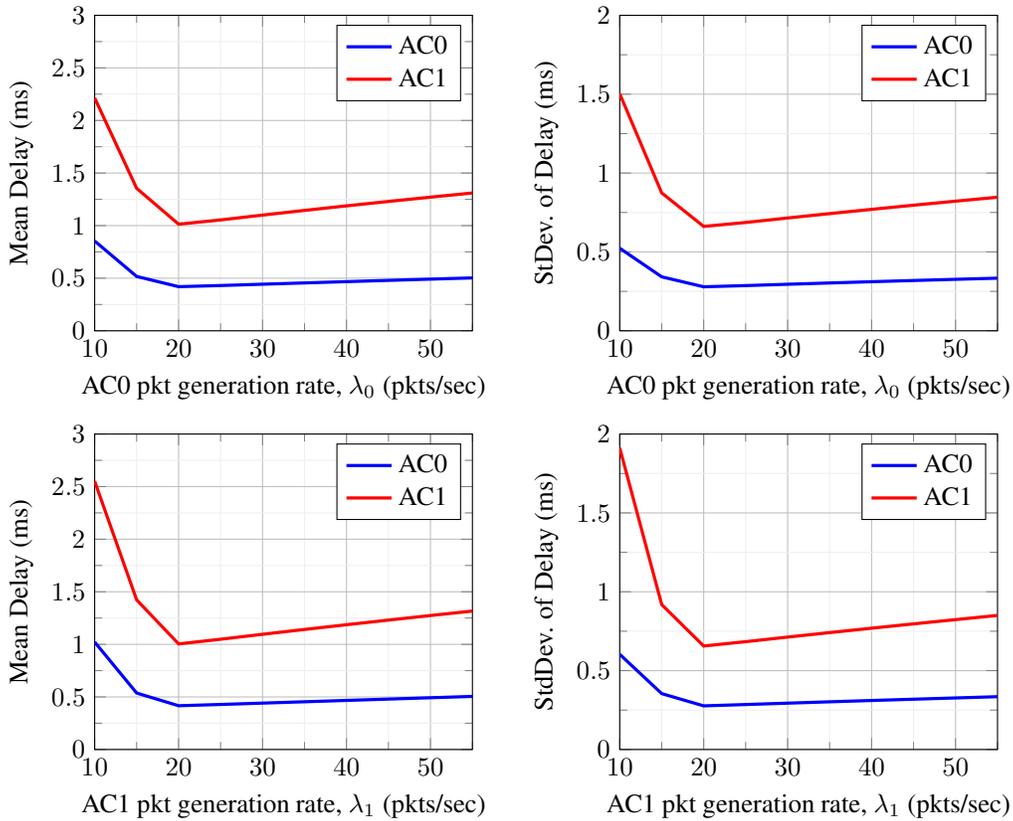

Using these computations, we analyzed the dependence of mean delay and standard deviation on the following parameters (i) number of transmitting vehicles, (ii) data rate, (iii) packet size, (iv) arrival rate of AC0 packets $\lambda_0$, and (v) arrival rate of AC1 packets $\lambda_1$. Each of these parameters was varied while all the other parameters were kept constant. The values of the parameters that were fixed are as mentioned in Table ~\ref{Tab:Tcr1}. The plots obtained for the mean and standard deviation of the MAC access delay for all the cases are as shown in Figures~\ref{fig:Mean_StDev_General1} and ~\ref{fig:Mean_StDev_General2}. 

As the number of transmitting vehicles increases, the MAC access delay increases, as seen in Figure~\ref{fig:Mean_StDev_General1}(a). This is indeed expected as increase in the number of vehicles causes an increase in the time spent by each transmitting vehicle contending for the channel. 
Figure~\ref{fig:Mean_StDev_General1}(b) shows that the MAC access delay reduces with increase in data rate, which is quite intuitive. Further, the MAC access delay is also seen to increase with increase in packet size, in Figure.~\ref{fig:Mean_StDev_General1}(c). 
A similar dependence is seen on the packet generation rates $\lambda_0$ and $\lambda_1$, see Figure~\ref{fig:Mean_StDev_General2}. For each access category, as the rate of packet generation increases, the corresponding queue in the MAC layer builds, thereby causing an increase in the time spent by the packets in the MAC layer. 

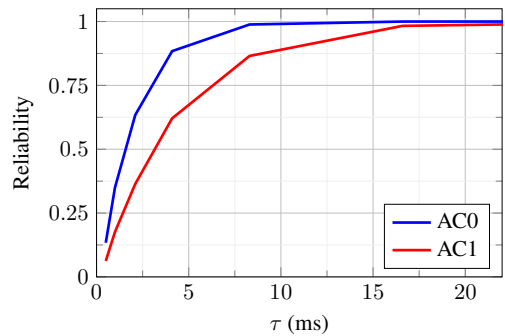
\begin{figure}[tbh]
\centering
\scalebox{0.85}{
\begin{tikzpicture}
\begin{axis}[
yticklabel style={
/pgf/number format/fixed,
/pgf/number format/precision=2,
/pgf/number format/fixed zerofill,
/pgf/number format/fixed relative,
},
    every axis plot/.append style={thick},
    xlabel = {$\tau$ (ms)},
    ylabel = {Reliability},
    xmin = 0, xmax = 22,
    ymin = 0, ymax = 1.05,
    xtick distance = 5,
    ytick distance = 0.25,
    grid = both,
    minor tick num = 1,
    major grid style = {lightgray},
    minor grid style = {lightgray!25},
    width = 0.48\textwidth,
    height = 0.35\textwidth,
    legend cell align = {right},
    legend pos = south east
]
 
\addplot[blue,style=very thick]  table [x = {x}, y = {y}] {rel_tau.prn};
\addplot[red,style=very thick]  table [x = {x}, y = {z}] {rel_tau.prn};
\legend{
AC0 , 
AC1
}
\end{axis}
\end{tikzpicture}
}
\caption{Reliability of the 802.11bd MAC protocol as a function of time variable $\tau$ plotted using the shifted exponential distribution given by equation~\eqref{eq:reliability}.} 
\label{fig:reliability_general}
\end{figure}

We also plotted the reliability of the MAC protocol outlined in equation~\eqref{eq:reliability} as a function of the time variable $\tau$ in the range $[0,20]$ ms. This plot is presented in Figure~\ref{fig:reliability_general}. From this plot, one can observe that the high priority access category has a higher reliability than the low priority category. Further, the likelihood of the data packets being delivered within an interval of 10 ms appears to be low for both the access categories. This indicates that it would be desirable to study the reliability of the protocol in the context of specific safety-related time-critical applications. 

Among the various parameters considered above, some depend on the traffic scenario that the transmitting vehicles are subjected to. In particular, the number of transmitting vehicles depends on the traffic density on the road, and the rate of packet generation would depend on the nature of traffic movement. For example, in order to perform cooperative movement such as platooning and lane changing, the vehicles would have to send their position/velocity information to the surrounding vehicles, both periodically as well as upon occurrence of a safety-critical event. The frequency of these transmissions, and hence the packet generation rates, would depend on complexity of the movement, the velocity of the vehicles, and the safety margins. Further, QoS requirements of vehicular communication protocols are often defined in terms of the target application. For example, for vehicular platooning, a maximum delay of 10-500 ms and a reliability higher than 90\% is desirable~\cite{naik2019ieee}. 
Therefore, it is desirable to understand the performance of the 802.11bd MAC protocol as applicable to particular traffic settings. One such application, that of platooning in a heterogeneous traffic setting, is discussed in Section 4. 


\section{Comparison with IEEE 802.11p}
In this subsection, we briefly digress to compare the Markov chain model and the analysis presented above with that of the standard EDCA mechanism, which is seen in the 802.11p protocol, which is the immediate predecessor of  802.11bd. There are two main differences between them, which are discussed below. 

In IEEE 802.11p~\cite{sun2013analytical}, the external collision probability is computed as 
    \begin{equation}
        p_{o} = 1 - \bigg(1- \sum_{i=0}^3 \beta_i\bigg)^{N_{cs} -1},
    \end{equation}
    where $\beta_i$ is the transmit probability of ACi. 
    Note that this is the same as that of 802.11bd when the first packet sent has been correctly decoded. Since 802.11p sends only 1 packet, $p(Z=1) = 1$. However, in 802.11bd, the number of packets sent differs based on the preamble detection and data decoding. Overall, it implies that in 802.11p, we only consider the collision of 1 packet while in 802.11bd, we need to consider the collision of 4 packets.

    In deriving the PGF for transmission time, we see that in 802.11p~\cite{sun2013analytical}, there is only 1 packet that is being transmitted, and its transmission time ($T_{tr}$) is deterministic, so the PGF turns out to be $z^{T_{tr}}$. However, in 802.11bd, there can be up to 4 packets to send, hence the transmission times for each of those events was considered leading to the derivation of equation~\eqref{eq:PGF_TR}. The same idea goes for the following equations.

\section{Application to platooning in heterogeneous traffic}
We now consider a platoon of connected vehicles surrounded by human-driven unconnected motorised two wheelers (MTWs). The connected vehicles that are equipped with V2V communication and use the 802.11bd protocol with EDCA and repetitions. The connected vehicles are required to maintain the platoon despite the interruptions by 
MTWs. 

Recall that, of the four access categories available in the 802.11bd standard, the high priority AC0 and AC1 are reserved for safety-related messages while the others are for infotainment. In our study, we only consider AC0 and AC1. Of these, AC0 will handle messages pertaining to interruptions by the MTWs as they are critical to safety, while the routine messages, such as the position and the velocity of the cars in the neighbourhood, will be catered by AC1. Hence, the packets for AC1 are generated periodically at a pre-defined rate while for AC0, the rate of packet generation is event-based and would depend on the probability of an MTW interrupting the platoon. Below we describe the mathematical models required to capture the behaviour of the platoon and the interruption by the MTWs. 

\subsection{Car-following Model: Platooning}
We considered the Full Velocity Difference (FVD) and Modified Optimal Velocity (MOV) models for our analysis. FVD models a platoon of $N$ cars moving on an infinite highway by using relative velocities and headways between car $i$ and car $i-1$ as follows~\cite{yu2013full}
\begin{align}
    \dot{v}_1(t) &= \ddot{x}_0(t) + a(\dot{x}_0(t) - V(y_1(t-\tau)) - v_1(t-\tau)) - lv_1(t-\tau),\notag \\ 
    \dot{v}_k(t) &= a(V(y_{k-1}(t-\tau)) - V(y_k(t-\tau)) - v_k(t-\tau)) + l(v_{k-1}(t-\tau) - v_k(t-\tau)), \label{eq:platoon_model}\\
    \dot{y}_i(t) &= v_i(t),\notag
\end{align}
where $v_i$ and $y_i$ represent relative velocities and headways of $i^{th}$ car. The indices $i = 1,2,$\dots$,N$ and $k = 2,3,\cdots,N$. The model parameters $a$ and $l$ capture the sensitivity of the vehicle to change in headway. $x_0$(t), $\dot{x}_0$(t) and $\ddot{x}_0$(t) represent the position, velocity and the acceleration of lead vehicle respectively. The feedback delay $\tau$ is the overall delay which consists of the communication delay and the delays due to actuation and braking in the vehicles. It is assumed to be constant for all vehicles in our analysis. The function $V(\cdot)$ represents the Bando Optimal Velocity Function (BOVF) and is given by~\cite{bando1995dynamical}
\begin{equation}
    V(y) = V_0 \bigg( \tanh \left( \frac{y-y_m}{\Tilde{y}} \right) + \tanh \left( \frac{y_m}{\Tilde{y}} \right)\bigg),\label{eq:OVF}
\end{equation}
where V$_0$, y$_m$ and $\Tilde{y}$ are model parameters. FVD boils down to MOVM when $l = 0$.
\vspace{-5pt}
\subsection{Gap Acceptance Model: Interruptions by MTWs}
We use gap acceptance model described in~\cite{bhattacharyya2020analytical} to model the rate at which the AC0 packets are generated at each car. The probability that an MTW will accept the gap between $i^{th}$ and ${i-1}^{th}$ cars moving in a platoon is given as
\begin{equation}
    P_i = exp \left(\alpha + \gamma \frac{y^\ast}{\dot{x}_i} \left[ 1 + exp \left( \alpha + \gamma \frac{y^\ast}{\dot{x}_i} \right) \right]^{-1} \right),\label{eq:gap_acceptance_probability}
\end{equation}
where $y^\ast$ is the equilibrium headway of the platoon and $\dot{x}_i$ is the velocity of car $i$ in a platoon. Since velocities of all cars remain constant, $P_i$ will remain fixed for all cars. Here, $\alpha$ and $\gamma$ are model parameters and it depends on the vehicle being considered in analysis. In our case, $\alpha = -1.933$ and $\gamma = 0.652$ as mentioned in \cite{bhattacharyya2020analytical}.

The rate of packet generation due to interruptions by MTW can now be modelled using equation~\eqref{eq:gap_acceptance_probability}. Since with increase in probability of such interruptions, the rate of interruptions $\lambda_0$ is also increasing. Hence it is intuitive to model the rate as a non-decreasing function of $P_i$. We use the following two non-decreasing functions in our analysis
\begin{itemize}
    \item Linear: $\lambda_0$ = $\eta\,P_i$ 
    \item Logarithmic: $\lambda_0$ = 1 - $\eta\log(1-P_i)$
\end{itemize}
where $\eta$ is a parameter.

\subsection{Non-oscillatory convergence of platoon headway}
To ensure safe platooning, we should satisfy the following three conditions. First, all vehicles should have the same constant headway. Second, their relative velocities should converge to 0. Lastly, the convergence should be non-oscillatory so as to ensure smooth and jerk-free movement. We now find a bound on $\tau$ that can satisfy the above three conditions. For this, we recapitulate the analysis outlined in~\cite{Gayathree802.11p}

The equilibrium headway $y^* = V^{-1} (\dot{x})$ and relative velocity of $i^{th}$ car $v^*_i = 0$. Let there be a perturbation $u_i(t) = y_i(t) - y^\ast$. Taking Taylor series expansion about $(y^\ast,0)$ gives the following linear approximation of model~\eqref{eq:platoon_model} 
\begin{align}
    \dot{v}_i(t) =&\, a V'(y^*) u_1(t-\tau) - (a+l)v_1(t-\tau),\notag \\
    \dot{v}_k(t) =&\, aV'(y^*) (u_{k-1}(t-\tau) - u_k(t-\tau)) - av_k(t-\tau) + l(v_{k-1}(t-\tau) - v_k(t-\tau)),\label{eq:platoon_model_linear} \\
    \dot{u}_{i}(t) =&\, v_i(t).\notag
\end{align}
Here $V'(y^\ast$) is the first derivative of OVF evaluated at $y^\ast$. Laplace Transform of system~\eqref{eq:platoon_model_linear} yields
\begin{equation}
    s^2 + (a+l)se^{-s\tau} + aV^{'}(y^*) e^{-s\tau} = 0. \label{eq:platoon_model_characteristic} 
\end{equation}
Substituting s = $\sigma + j\omega$ in equation~\eqref{eq:platoon_model_characteristic} and separating the real and imaginary parts yields the following two equations
\begin{align}
    \mathcal{A}e^{-\sigma \tau} cos(\omega \tau) + \mathcal{B} e^{-\sigma \tau} sin(\omega \tau) = \omega^2 - \sigma^2, \label{eq:platoon_model_char_real}\\
    \mathcal{B} e^{-\sigma \tau} cos(\omega \tau) - \mathcal{A} e^{-\sigma \tau} sin(\omega \tau) = -2\sigma \omega,\label{eq:platoon_model_char_imag}
\end{align}
where $\mathcal{A} = \sigma (a+l) + a V'(y^*)$ and $\mathcal{B} = (a+l)\omega$. Squaring and adding equations~\eqref{eq:platoon_model_char_real} and~\eqref{eq:platoon_model_char_imag} yields
\begin{equation}
    \mathcal{A}^2 + \mathcal{B}^2 = \left((\omega^2 - \sigma^2)^2 + 4\sigma^2 \omega^2\right)e^{2\sigma \tau}.\label{eq:sum_of_squares}
\end{equation}
For non-oscillatory convergence of the solutions of system~\eqref{eq:platoon_model_linear}, we require that the roots of the characteristic equation~\eqref{eq:platoon_model_characteristic} be purely real. 
For $s$ to be a real root, we require that $\omega = 0$ must be 0 and it should be a unique solution to equation~\eqref{eq:sum_of_squares}. We determine the necessary conditions and the sufficient conditions as follows. For necessary condition, put $\omega =0$ in equation~\eqref{eq:sum_of_squares} to obtain
\begin{equation}
    \mathcal{A}^2 = \sigma^4 e^{2\sigma \tau}.
\end{equation}
For the sufficient condition, substitute $\mathcal{A}^2$ in equation~\eqref{eq:sum_of_squares} to get 
\begin{align}
    \sigma^4 e^{2\sigma \tau} + (a+l)^2\omega^2 - (\omega^2 + \sigma^2)^2 e^{2\sigma \tau} &= 0,\label{eq:suff_1} \\
    \omega^2 (\omega^2 + 2\sigma^2 - (a+l)^2 e^{-2\sigma \tau}) &= 0. \label{eq:suff_2}
\end{align}
For $\omega =0$ to be a unique solution, we must have $ 2\sigma^2 e^{2\sigma \tau} = (a+l)^2. $
Thus, we must satisfy conditions~\eqref{eq:suff_1} and~\eqref{eq:suff_2} for the characteristic equation~\eqref{eq:platoon_model_characteristic} to have real roots. Solving it further, we get
\begin{equation}
    \sigma = \Tilde{d}(-2 \pm \sqrt{2}), 
\end{equation}
where $\Tilde{d} = aV'(y^*)/(a+l)$. Putting the value of $\sigma$ and $\omega = 0$ in equation~\eqref{eq:platoon_model_characteristic}, we get the value of $\tau_C$ as
\begin{equation}
    \tau_C = \frac{-1}{\title{d} (2 + \sqrt{2})} ln \left( \frac{-\Tilde{d}(a+l)(-2-\sqrt{2} - aV'(y^*))}{\Tilde{d}^2(-2 - \sqrt{2})^2} \right),\label{eq:tau_non_oscill}
\end{equation}
and we require $\tau \leq \tau_C$ for non-oscillatory convergence.

Note that $\tau_C$ is the upper bound on feedback delay, which consists of communication delay and mechanical delay, for non-oscillatory convergence of the headways of the vehicles in the platoon. It is shown in~\cite{beregi2021connectivity} that mechanical delays usually comprise up to 85-90\% of the total feedback delay. Thus, we consider the bound on communication delay as 10\% of $\tau_C$ as follows 
\begin{equation}
    \tau_{cr} = \kappa \tau_C,\label{eq:tau_comm_non_oscill}
\end{equation}
where $\kappa \sim \mathcal{N}(0.1, 0.01)$.

\begin{figure}[tbh]
\centering
    \begin{tabular}[c]{c c}
\begin{tikzpicture}
\begin{axis}[
yticklabel style={
/pgf/number format/fixed,
/pgf/number format/precision=2,
/pgf/number format/fixed zerofill,
/pgf/number format/fixed relative,
},
    every axis plot/.append style={thick},
    xlabel = {Headway (m)},
    ylabel = {Mean Delay (ms)},
    xmin = 2, xmax = 10,
    ymin = 0, ymax = 4.5,
    xtick distance = 2,
    ytick distance = 1,
    grid = both,
    minor tick num = 1,
    major grid style = {lightgray},
    minor grid style = {lightgray!25},
    width = 0.4\textwidth,
    height = 0.35\textwidth,
    legend cell align = {right},
    legend pos = north east
]
 
\addplot[blue,style=very thick]  table [x = {x}, y = {y}] {delay_headway.prn};
\addplot[red,style=very thick]  table [x = {x}, y = {a}] {delay_headway.prn};
\addplot[blue, only marks,mark=square,style= thick]  table [x = {x}, y = {y}, meta = {label}] {delay_headway_sim.prn};
\addplot[red, only marks,mark=*,style = thick]  table [x = {x}, y = {a}, meta = {label}] {delay_headway_sim.prn};
\legend{
AC0 Analy., 
AC1 Analy.,
AC0 Sim.,
AC1 Sim.}
\end{axis}
\end{tikzpicture}
&
\begin{tikzpicture}
\begin{axis}[
yticklabel style={
/pgf/number format/fixed,
/pgf/number format/precision=2,
/pgf/number format/fixed zerofill,
/pgf/number format/fixed relative,
},
    every axis plot/.append style={thick},
    xlabel = {Headway (m)},
    ylabel = {StdDev. of Delay (ms)},
    xmin = 2, xmax = 10,
    ymin = 0, ymax = 4.5,
    xtick distance = 2,
    ytick distance = 1,
    grid = both,
    minor tick num = 1,
    major grid style = {lightgray},
    minor grid style = {lightgray!25},
    width = 0.4\textwidth,
    height = 0.35\textwidth,
    legend cell align = {right},
    legend pos = north east
]
 
\addplot[blue,style=very thick]  table [x = {x}, y = {z}] {delay_headway.prn};
\addplot[red,style=very thick]  table [x = {x}, y = {b}] {delay_headway.prn};
\addplot[blue, only marks,mark=square,style= thick]  table [x = {x}, y = {z}, meta = {label}] {delay_headway_sim.prn};
\addplot[red, only marks,mark=*,style = thick]  table [x = {x}, y = {b}, meta = {label}] {delay_headway_sim.prn};
\legend{
AC0 Analy., 
AC1 Analy.,
AC0 Sim.,
AC1 Sim.}
\end{axis}
\end{tikzpicture}
\\
\multicolumn{2}{c}{(a) Linear function}
\\
\begin{tikzpicture}
\begin{axis}[
yticklabel style={
/pgf/number format/fixed,
/pgf/number format/precision=2,
/pgf/number format/fixed zerofill,
/pgf/number format/fixed relative,
},
    every axis plot/.append style={thick},
    xlabel = {Headway (m)},
    ylabel = {Mean Delay (ms)},
    xmin = 2, xmax = 10,
    ymin = 0, ymax = 4,
    xtick distance = 2,
    ytick distance = 1,
    grid = both,
    minor tick num = 1,
    major grid style = {lightgray},
    minor grid style = {lightgray!25},
    width = 0.4\textwidth,
    height = 0.35\textwidth,
    legend cell align = {right},
    legend pos = north east
]
 
\addplot[blue,style=very thick]  table [x = {x}, y = {y}] {delay_headway_log.prn};
\addplot[red,style=very thick]  table [x = {x}, y = {a}] {delay_headway_log.prn};
\addplot[blue, only marks,mark=square,style= thick]  table [x = {x}, y = {y}] {delay_headway_log_sim.prn};
\addplot[red, only marks,mark=*,style = thick]  table [x = {x}, y = {a}] {delay_headway_log_sim.prn};
\legend{
AC0 Analy., 
AC1 Analy.,
AC0 Sim.,
AC1 Sim.}
\end{axis}
\end{tikzpicture}
&
\begin{tikzpicture}
\begin{axis}[
yticklabel style={
/pgf/number format/fixed,
/pgf/number format/precision=2,
/pgf/number format/fixed zerofill,
/pgf/number format/fixed relative,
},
    every axis plot/.append style={thick},
    xlabel = {Headway (m)},
    ylabel = {StdDev. of Delay (ms)},
    xmin = 2, xmax = 10,
    ymin = 0, ymax = 2.5,
    xtick distance = 2,
    ytick distance = 0.5,
    grid = both,
    minor tick num = 1,
    major grid style = {lightgray},
    minor grid style = {lightgray!25},
    width = 0.4\textwidth,
    height = 0.35\textwidth,
    legend cell align = {right},
    legend pos = north east
]
 
\addplot[blue,style=very thick]  table [x = {x}, y = {z}] {delay_headway_log.prn};
\addplot[red,style=very thick]  table [x = {x}, y = {b}] {delay_headway_log.prn};
\addplot[blue, only marks,mark=square,style= thick]  table [x = {x}, y = {z}] {delay_headway_log_sim.prn};
\addplot[red, only marks,mark=*,style = thick]  table [x = {x}, y = {b}] {delay_headway_log_sim.prn};
\legend{
AC0 Analy., 
AC1 Analy.,
AC0 Sim.,
AC1 Sim.}
\end{axis}
\end{tikzpicture}
\\
\multicolumn{2}{c}{(b) Logarithmic function}
    \end{tabular}
    \caption{Mean and standard deviation of the MAC access delay as a function of the platoon headway $(y^\ast)$. The MAC acess delay is seen to be inversely related to the headway.}
    \label{fig:delay_characterstics_platoon}
\end{figure}
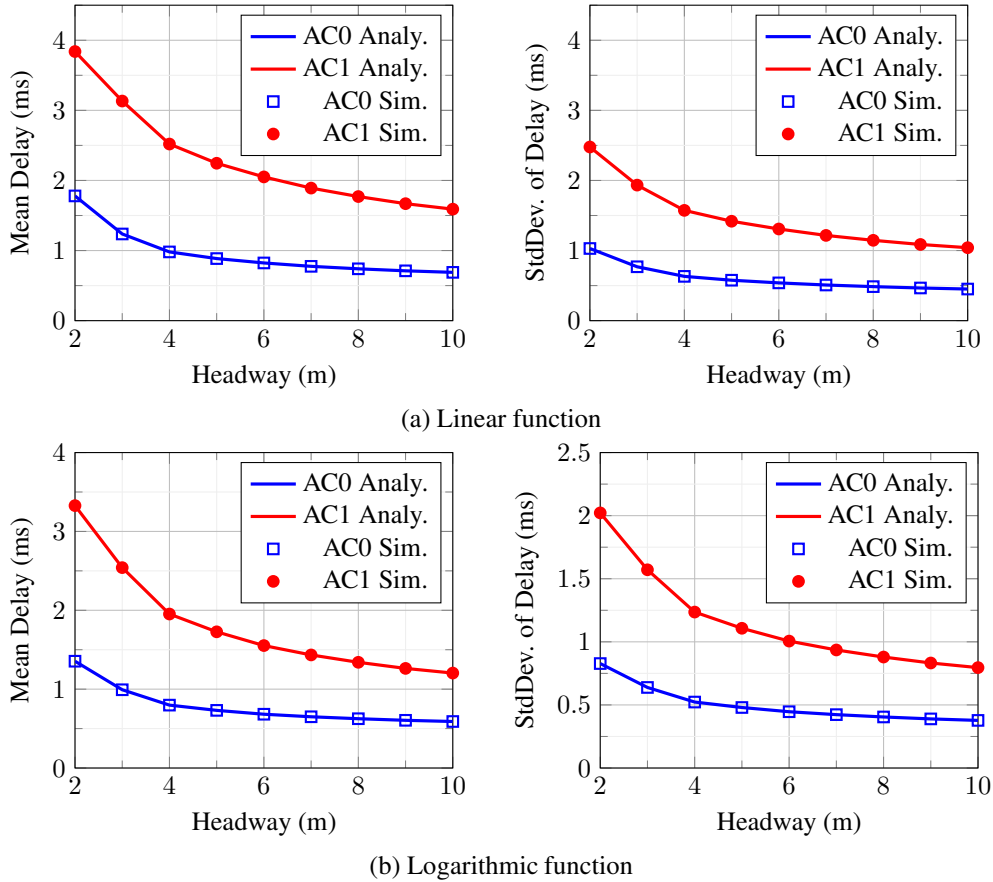

\subsection{Computations and simulations}
From the aforementioned analysis, we compute two parameters: (i) the mean rate of generation of safety-critical packets AC0 using the gap-acceptance model outlined in Section~4.2, (ii) the critical communication delay for non-oscillatory convergence of the platoon using equation~\eqref{eq:tau_comm_non_oscill}.  Note that both these parameters are a function of the equilibrium headway $y^\ast$ of the platoon. 

\begin{figure}[tbh]
\centering
\begin{tabular}[c]{c c}
\begin{tikzpicture}
\begin{axis}[
yticklabel style={
/pgf/number format/fixed,
/pgf/number format/precision=2,
/pgf/number format/fixed zerofill,
/pgf/number format/fixed relative,
},
    every axis plot/.append style={thick},
    ylabel = {Reliability},
    xlabel = {Headway (m)},
    xmin = 2, xmax = 10,
    ymin = 0.4, ymax = 1.05,
    xtick distance = 2,
    ytick distance = 0.2,
    grid = both,
    minor tick num = 1,
    major grid style = {lightgray},
    minor grid style = {lightgray!25},
    width = 0.4\textwidth,
    height = 0.35\textwidth,
    legend cell align = {right},
    legend pos = south west
]
\addplot[blue,style=very thick]  table [x = {x}, y = {c}] {rel_headway.prn};
\addplot[red,style=very thick]  table [x = {x}, y = {d}] {rel_headway.prn};
\legend{
AC0 , 
AC1
}
\end{axis}
\end{tikzpicture}
&
\begin{tikzpicture}
\begin{axis}[
yticklabel style={
/pgf/number format/fixed,
/pgf/number format/precision=2,
/pgf/number format/fixed zerofill,
/pgf/number format/fixed relative,
},
    every axis plot/.append style={thick},
    ylabel = {Reliability},
    xlabel = {Headway (m)},
    xmin = 2, xmax = 10,
    ymin = 0.4, ymax = 1.05,
    xtick distance = 2,
    ytick distance = 0.2,
    grid = both,
    minor tick num = 1,
    major grid style = {lightgray},
    minor grid style = {lightgray!25},
    width = 0.4\textwidth,
    height = 0.35\textwidth,
    legend cell align = {right},
    legend pos = south west
]
 
\addplot[blue,style=very thick]  table [x = {x}, y = {e}] {rel_headway.prn};
\addplot[red,style=very thick]  table [x = {x}, y = {f}] {rel_headway.prn};
\legend{
AC0 , 
AC1
}
\end{axis}
\end{tikzpicture}\\
(i) Linear function & (ii) Logarithmic function\\
\end{tabular}
\caption{Reliability of 802.11bd MAC protocol for a platoon following the Full Velocity Difference (FVD) model subject to interruptions by human-driven MTWs.}
\label{fig:reliability_platoon}
\end{figure}
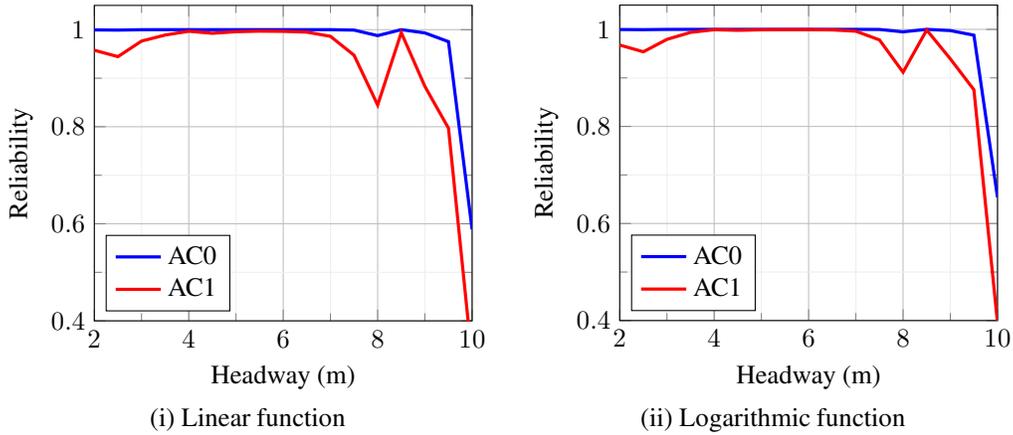

We first conducted computations to analyse the characteristics of the MAC access delay with respect to the platoon headway. For this, we varied the headway in the range $y^\ast \in [2,10]$ m, and for each value computed $\lambda_0$, the mean generation rate of AC0 packets. Further, we fixed $\lambda_1$, the rate of generation of AC1 packets as $30$ pkst/sec. With these values, we computed the mean and standard deviation of the MAC access delay, similar to computations described in Section~2.5. We also simulated EDCA algorithm with repetitions as in the IEEE 802.11bd MAC protocol using virtual queues for each access category with enqueue and dequeue times as per the contention algorithm, for the same parameter values. The mean and standard deviation of the delay obtained through these simulations are juxtaposed with the computational results. The plots are presented in Figure~\ref{fig:delay_characterstics_platoon}. The computational results are seen to match with the simulations. Further, the delay is seen to reduce with increase in the platoon headway. This is similar to the observation of delay characteristics with respect to number of vehicles, in Figure~\ref{fig:Mean_StDev_General1}. Indeed, as the headway increases, the number of vehicles within the contention range decreases, which causes a reduction in the time that each packet spends contending for the channel. Next, we computed the reliability of the protocol using equation~\eqref{eq:reliability} and value of $\tau_{cr}$ computed using equation~\eqref{eq:tau_comm_non_oscill}. These plots are presented in Figure~\ref{fig:reliability_platoon}. From these plots, we observe that the reliability of the protocol reduces for values of headway higher than $8$ m. This is a result of reduced critical delay for larger headways. This preliminary analysis shows that the 802.11bd MAC may fail to transmit safety-critical packets as per the delay requirements of platooning in heterogeneous traffic setup. All the computations and simulations were performed using the scientific computing software MATLAB.

\section{Conclusion}
In this article, we presented an analytical study of IEEE 802.11bd MAC protocol that uses EDCA with repetitions. We used a Markov-chain model for the backoff mechanism of the protocol and then used probability generating functions to compute the mean and standard deviation of the MAC access delay. We also derived an expression for the protocol reliability, which is the likelihood that a packet can be delivered within a critical time. We then used computations to understand the dependence of the MAC access delay on various parameters such as number of transmitting vehicles, rate of packet generation and other network parameters. As these parameters are dependent on specific applications and the safety-critical events thereof. 

Motivated by this, we investigated the delay performance of the protocol in the specific context of a platoon of connected cars interrupted by human-driven motorized two-wheelers. Models such as Full Velocity Difference (FVD) and Modified Optimal Velocity (MOV) were used for modelling platoons and Gap-Acceptance model was used to model the rate of MTWs interrupting the platoon. We used these models to compute the rate of generation of safety-critical packets AC0 and the critical communication delay for jerk-free platooning. Using these parameters, we recomputed the delay characteristics and the protocol reliability. It is seen that the delay performance of the protocol is indeed dependent on the traffic settings which influence the network parameters. Our study suggests that it would be desirable to co-design channel access protocols for vehicular communication with specific application of cooperative movement.  

\bibliographystyle{unsrt}  
\bibliography{main} 

\end{document}